\documentclass[prd,superscriptaddress,showpacs,twocolumn]{revtex4}

\usepackage{amsmath}
\usepackage{slashed}
\usepackage{feynmp}

\renewcommand{\Re}{\mathop{\rm Re}\nolimits}
\renewcommand{\Im}{\mathop{\rm Im}\nolimits}
\newcommand{\Tr}{\mathop{\rm Tr}\nolimits}
\newcommand{\bra}[1]{\langle#1\vert}
\newcommand{\ket}[1]{\vert#1\rangle}

\begin{document}
\begin{fmffile}{fig}

\title{Dynamical fermion mass generation by a~strong Yukawa interaction}
\author{Tom\'a\v{s} Brauner}
\email{brauner@ujf.cas.cz}
\affiliation{Department of Theoretical Physics, Nuclear Physics Institute, \v Re\v z, Czech Republic}
\affiliation{Faculty of Mathematics and Physics, Charles University, Prague, Czech Republic}
\author{Ji\v{r}\'{\i} Ho\v{s}ek}
\affiliation{Department of Theoretical Physics, Nuclear Physics Institute, \v Re\v z, Czech Republic}

\begin{abstract}
We consider a~model with global Abelian chiral symmetry of two massless fermion
fields interacting with a~complex massive scalar field. We argue that the
Schwinger--Dyson equations for the fermion and boson propagators admit
ultraviolet-finite chiral-symmetry-breaking solutions provided the Yukawa
couplings are large enough. The fermions acquire masses and the elementary
excitations of the complex scalar field are the two real spin-zero particles
with different masses. As a~necessary consequence of the dynamical chiral
symmetry breakdown both in the fermion and scalar sectors, one massless
pseudoscalar Nambu--Goldstone boson appears in the spectrum as a~collective
excitation of both the fermion and the boson fields. Its effective couplings to
the fermion and boson fields are calculable.
\end{abstract}

\pacs{11.30.Qc}
\maketitle

\section{Introduction}
One of the major challenges to current high energy physics is to understand the
origin of particle masses \cite{Quigg:2005hr}. All particle interactions except
gravity are, up to energies accessible in to-date experiments, successfully
described by the Standard model. There is therefore no doubt that the Standard
model is the correct \emph{effective} theory of particle interactions in the
energy range so far explored.

It is well known that the detailed underlying physics manifests itself in the
effective theory through the effective coupling constants so that when looking
for new physics, we should try to reveal the origin of the parameters in the
effective Lagrangian.

In the Standard model, the part of the Lagrangian describing the particle
interactions is beautifully constrained by the gauge invariance principle. The
matter part, however, seems rather ugly. The fermion masses cannot be
introduced directly as they are prohibited by the gauge $SU(2)_L\times U(1)_Y$
symmetry, whose chiral structure is in turn necessary to provide an explanation
for parity violation in weak interactions.

The masses therefore have to be generated by means of spontaneous symmetry
breaking. This is achieved by introducing the scalar Higgs field and properly
adjusting its potential so that it develops a~non-zero vacuum expectation value
at tree level. However, the dynamical origin of the `wrong sign' of the Higgs
mass squared, which is responsible for the tree-level condensation, remains
unclear. Moreover, the fermion masses eventually stem from the Yukawa
interaction with the Higgs, and hence there are as many different couplings as
there are particle species. It would certainly be desirable to understand the
particle masses as the consequence of some, yet unknown, quantum dynamics.

We would like to argue that dynamical mass generation is possible by means of
the Yukawa interaction itself, without ever having to change the sign of the
Higgs mass squared. This would open a~new possibility of the existence of an
elementary massive scalar field in the Standard model Lagrangian. It should be
heavy enough so that current experimental bounds are met, and also due to
naturalness arguments, according to which its mass should be shifted by
radiative corrections up to the scale of new physics.

Our ambitions in the present paper are much more modest than to cure the
Standard model from its disease. We disregard the otherwise phenomenologically
very important issues such as the hierarchy of particle families and CP
violation, and concentrate on a~simple Abelian model to show that spontaneous
chiral symmetry breaking by the Yukawa interaction is viable. We believe that
the central idea of the symmetry-breaking mechanism can thus be displayed more
transparently. The implementation of this idea to the Standard model
phenomenology is deferred to future work.

The plan of the paper is as follows. In the next section we introduce our model
and investigate, at a~rather elementary and pictorial level, the consequences
of the assumed spontaneous generation of the fermion mass. We do so for the
reader's convenience and to emphasize the robustness of the main idea of the
paper. In particular we show that the fermion mass induces an anomalous
symmetry-breaking two-point Green's function of the scalar. The spectrum of the
system then contains two real spinless particles coupled to the original
complex field, and with their masses split.

In the next part of the text a~matrix formalism is developed which allows us to
treat both the symmetry-preserving and the symmetry-breaking Green's functions
on the same footing. We write down the one-loop Schwinger--Dyson equations and
exploit the underlying symmetry by means of the Ward identities. Still
\emph{assuming} the spontaneous symmetry breaking to occur in the ground state,
we show how the Nambu--Goldstone boson arises and examine its properties.

In the last part we demonstrate that, under reasonable simplifying assumptions
a~symmetry-breaking solution to the Schwinger--Dyson equations actually does
exist, thus establishing a~firm ground for the preceding heuristic arguments.
We conclude the paper with a~discussion of our results and the future
perspective.

\section{Preliminary considerations}
\label{subsec_preliminary}
Our model is defined by the Lagrangian,
\begin{multline}
{\cal L}=\sum_{j=1,2}\left(\bar\psi_{jL}i\slashed{\partial}\psi_{jL}+
\bar\psi_{jR}i\slashed{\partial}\psi_{jR}\right)+\\
+\partial_{\mu}\phi^{\dagger}\partial^{\mu}\phi-M^2\phi^{\dagger}\phi
-\frac12\lambda(\phi^{\dagger}\phi)^2+\\
\shoveright{+y_1\bar\psi_{1L}\psi_{1R}\phi+y_1\bar\psi_{1R}\psi_{1L}\phi^{\dagger}+}\\
+y_2\bar\psi_{2R}\psi_{2L}\phi+y_2\bar\psi_{2L}\psi_{2R}\phi^{\dagger}.
\label{lagrangian}
\end{multline}
The Yukawa couplings $y_{1,2}$ are real without lack of generality. Another
remark is in order here. In view of the future application of our idea on the
electroweak symmetry breaking, it is necessary that the global symmetry to be
spontaneously broken is amenable to gauging. With just a~single fermion species
there would be an axial anomaly present. While anomaly cancelation is automatic
in the Standard model due to its particle content, here we have to introduce
two fermions with opposite axial charges to remove the anomaly in the Abelian
axial current. It should be clear, however, that this minor technical
complication does not alter at all the underlying idea.

The Lagrangian Eq. \eqref{lagrangian} enjoys a~global Abelian $U(1)_{V1}\times
U(1)_{V2}\times U(1)_A$ symmetry. The two vector $U(1)$'s are associated with
separate conservation of the numbers of fermions of the first and second type.
The corresponding Noether currents are the well known
\begin{equation}
\begin{split}
j_{V1}^{\mu}&=\bar\psi_1\gamma^{\mu}\psi_1,\\
j_{V2}^{\mu}&=\bar\psi_2\gamma^{\mu}\psi_2.
\end{split}
\label{Noether_vector}
\end{equation}

The axial $U(1)$ transformations of the fermions are tied by the Yukawa
coupling to the scalar field. For the Lagrangian to be invariant, it is
necessary that the fields transform as
\begin{equation}
\begin{split}
\psi_1&\to e^{+i\theta\gamma_5}\psi_1,\\
\psi_2&\to e^{-i\theta\gamma_5}\psi_2,\\
\phi&\to e^{-2i\theta}\phi.
\end{split}
\label{axial_transform}
\end{equation}
The axial Noether current has the following form,
\begin{equation}
j_A^{\mu}=\bar\psi_1\gamma^{\mu}\gamma_5\psi_1-\bar\psi_2\gamma^{\mu}\gamma_5\psi_2+
2i\left[(\partial^{\mu}\phi)^{\dagger}\phi-\phi^{\dagger}\partial^{\mu}\phi\right].
\label{Noether_axial}
\end{equation}

Now the standard tree-level mechanism of spontaneous symmetry breaking
corresponds to $M^2<0$ in Eq. \eqref{lagrangian}. Then the scalar field,
$\phi=\frac1{\sqrt2}(\phi_1+i\phi_2)$, develops a~non-zero ground-state
expectation value, which is conveniently chosen to be real,
$v\equiv\langle\phi_1\rangle_0=(-2M^2/\lambda)^{1/2}$. Consequently, the
fermions acquire the masses $m_{1,2}=\frac1{\sqrt2}vy_{1,2}$, $\phi_2$ becomes
the massless Nambu--Goldstone boson, and $\phi_1'=\phi_1-v$ becomes a~massive
real scalar particle with mass $M_1=\sqrt{-2M^2}$. The Goldstone boson $\phi_2$
interacts with the fermions by the Yukawa coupling
$\frac{m_1}v\bar\psi_1i\gamma_5\psi_1\phi_2-\frac{m_2}v\bar\psi_2i\gamma_5\psi_2\phi_2$.

From now on we set $M^2>0$ and investigate the model of Eq. \eqref{lagrangian}
with respect to the possibility of spontaneous symmetry breaking. The scalar
field now possesses an ordinary mass term, and there is therefore no
condensation and no symmetry breaking at tree level. Rather, if the symmetry is
broken in the ground state, it must be a~result of the quantum dynamics of the
system.

As spontaneous symmetry breaking is a~non-perturbative phenomenon, it cannot be
achieved at any finite order of perturbation theory. The non-perturbative
method we employ here is to look for self-consistent symmetry-breaking
solutions to the Schwinger--Dyson equations. We thus temporarily \emph{assume}
that there is a~solution for which the full fermion propagators have
non-vanishing chirality-changing parts, which means non-zero masses of the
fermions.

\begin{figure}
$$
\parbox{30\unitlength}{%
\begin{fmfgraph*}(30,15)
\fmfset{arrow_len}{0.08w}
\fmfleft{l}
\fmfright{r}
\fmf{scalar,label=$\phi$,l.si=left,tension=4}{vl,l}
\fmf{scalar,label=$\phi$,l.si=right,tension=4}{vr,r}
\fmf{phantom,right}{vr,vl}
\fmf{phantom,right}{vl,vr}
\fmffreeze
\fmfipath{p[]}
\fmfiset{p1}{vpath(__vr,__vl)}
\fmfiset{p2}{vpath(__vl,__vr)}
\fmfiv{d.sh=circle,d.si=0.1w,d.fi=full}{point length(p1)/2 of p1}
\fmfiv{d.sh=circle,d.si=0.1w,d.fi=full}{point length(p2)/2 of p2}
\fmfi{fermion,label=$\psi_{1R}$,l.si=right}{subpath(0,length(p1)/2) of p1}
\fmfi{fermion,label=$\psi_{1L}$,l.si=right}{subpath(length(p1)/2,length(p1)) of p1}
\fmfi{fermion,label=$\psi_{1R}$,l.si=right}{subpath(0,length(p2)/2) of p2}
\fmfi{fermion,label=$\psi_{1L}$,l.si=right}{subpath(length(p2)/2,length(p2)) of p2}
\end{fmfgraph*}}
\,+\,
\parbox{30\unitlength}{%
\begin{fmfgraph*}(30,15)
\fmfset{arrow_len}{0.08w}
\fmfleft{l}
\fmfright{r}
\fmf{scalar,label=$\phi$,l.si=left,tension=4}{vl,l}
\fmf{scalar,label=$\phi$,l.si=right,tension=4}{vr,r}
\fmf{phantom,right}{vr,vl}
\fmf{phantom,right}{vl,vr}
\fmffreeze
\fmfipath{p[]}
\fmfiset{p1}{vpath(__vr,__vl)}
\fmfiset{p2}{vpath(__vl,__vr)}
\fmfiv{d.sh=circle,d.si=0.1w,d.fi=full}{point length(p1)/2 of p1}
\fmfiv{d.sh=circle,d.si=0.1w,d.fi=full}{point length(p2)/2 of p2}
\fmfi{fermion,label=$\psi_{2L}$,l.si=right}{subpath(0,length(p1)/2) of p1}
\fmfi{fermion,label=$\psi_{2R}$,l.si=right}{subpath(length(p1)/2,length(p1)) of p1}
\fmfi{fermion,label=$\psi_{2L}$,l.si=right}{subpath(0,length(p2)/2) of p2}
\fmfi{fermion,label=$\psi_{2R}$,l.si=right}{subpath(length(p2)/2,length(p2)) of p2}
\end{fmfgraph*}}
$$
\caption{Mixing in scalar sector induced by the fermion mass terms. The full
blobs denote the chirality-changing part of the complete fermion propagator.}
\label{Fig:phi-phi_mixing}
\end{figure}
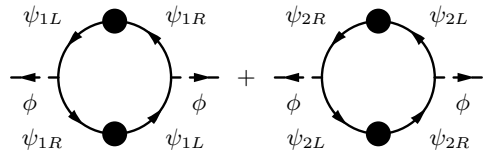

Now let us observe that such a~solution induces mixing of $\phi$ and
$\phi^{\dagger}$ or, in other words, non-zero correlation function
$\langle\phi\phi\rangle$, see Fig. \ref{Fig:phi-phi_mixing}. This could have
been expected as the field $\phi$ couples to the axial part of the
$U(1)_{V1}\times U(1)_{V2}\times U(1)_A$ symmetry, and once this is broken
nothing prevents $\phi$ and $\phi^{\dagger}$ from mixing.

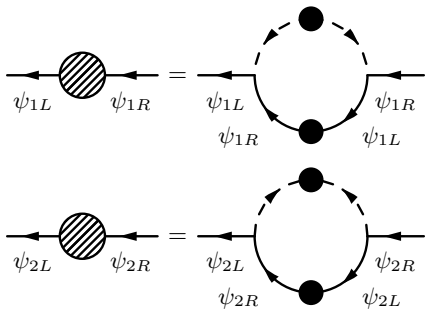
\begin{figure}
\begin{align*}
\parbox{20\unitlength}{%
\begin{fmfgraph*}(20,15)
\fmfset{arrow_len}{0.12w}
\fmfleft{l}
\fmfright{r}
\fmf{fermion,label=$\psi_{1L}$,l.si=left}{v,l}
\fmf{fermion,label=$\psi_{1R}$,l.si=left}{r,v}
\fmfv{d.sh=circle,d.si=0.3w,d.fi=shaded}{v}
\end{fmfgraph*}}
\,&=\,
\parbox{30\unitlength}{%
\begin{fmfgraph*}(30,20)
\fmfset{arrow_len}{0.08w}
\fmfleft{l}
\fmfright{r}
\fmf{fermion,tension=4,label=$\psi_{1L}$,l.si=left}{vl,l}
\fmf{fermion,tension=4,label=$\psi_{1R}$,l.si=left}{r,vr}
\fmf{phantom,right}{vr,vl}
\fmf{phantom,right}{vl,vr}
\fmffreeze
\fmfipath{p[]}
\fmfiset{p1}{vpath(__vr,__vl)}
\fmfiset{p2}{vpath(__vl,__vr)}
\fmfiv{d.sh=circle,d.si=0.1w,d.fi=full}{point length(p1)/2 of p1}
\fmfiv{d.sh=circle,d.si=0.1w,d.fi=full}{point length(p2)/2 of p2}
\fmfi{scalar}{subpath(length(p1)/2,0) of p1}
\fmfi{scalar}{subpath(length(p1)/2,length(p1)) of p1}
\fmfi{fermion,label=$\psi_{1R}$,l.si=left}{subpath(length(p2)/2,0) of p2}
\fmfi{fermion,label=$\psi_{1L}$,l.si=left}{subpath(length(p2),length(p2)/2) of p2}
\end{fmfgraph*}}\\
\parbox{20\unitlength}{%
\begin{fmfgraph*}(20,15)
\fmfset{arrow_len}{0.12w}
\fmfleft{l}
\fmfright{r}
\fmf{fermion,label=$\psi_{2L}$,l.si=left}{v,l}
\fmf{fermion,label=$\psi_{2R}$,l.si=left}{r,v}
\fmfv{d.sh=circle,d.si=0.3w,d.fi=shaded}{v}
\end{fmfgraph*}}
\,&=\,
\parbox{30\unitlength}{%
\begin{fmfgraph*}(30,20)
\fmfset{arrow_len}{0.08w}
\fmfleft{l}
\fmfright{r}
\fmf{fermion,tension=4,label=$\psi_{2L}$,l.si=left}{vl,l}
\fmf{fermion,tension=4,label=$\psi_{2R}$,l.si=left}{r,vr}
\fmf{phantom,right}{vr,vl}
\fmf{phantom,right}{vl,vr}
\fmffreeze
\fmfipath{p[]}
\fmfiset{p1}{vpath(__vr,__vl)}
\fmfiset{p2}{vpath(__vl,__vr)}
\fmfiv{d.sh=circle,d.si=0.1w,d.fi=full}{point length(p1)/2 of p1}
\fmfiv{d.sh=circle,d.si=0.1w,d.fi=full}{point length(p2)/2 of p2}
\fmfi{scalar}{subpath(0,length(p1)/2) of p1}
\fmfi{scalar}{subpath(length(p1),length(p1)/2) of p1}
\fmfi{fermion,label=$\psi_{2R}$,l.si=left}{subpath(length(p2)/2,0) of p2}
\fmfi{fermion,label=$\psi_{2L}$,l.si=left}{subpath(length(p2),length(p2)/2) of p2}
\end{fmfgraph*}}
\end{align*}
\caption{Chirality changing fermion proper self-energies induced by the scalar
mixing.}
\label{Fig:fermion_propagator}
\end{figure}

The second observation is that the `anomalous' two-point scalar Green's
function $\langle\phi\phi\rangle$ in turn enters the one-loop Schwinger--Dyson
equations for the fermion propagators, see Fig. \ref{Fig:fermion_propagator}.
The set of Schwinger--Dyson equations for the fermion and scalar propagators
are thus intrinsically coupled and must be solved simultaneously if a~symmetry
breaking solution is to be found.

Before we switch to a~formal description to come in the next section, we would
like to give a~more physical picture of what is going on here. Let us for
simplicity suppose that the one-particle-irreducible part of the anomalous
scalar two-point function $\langle\phi\phi\rangle$ is momentum independent and
equal to $-i\mu^2$. The spectrum in the scalar sector is then determined by the
quadratic part of the renormalized Lagrangian,
\begin{equation}
{\cal L}^{(0)}_{\text{scalar}}=\partial_{\mu}\phi^{\dagger}\partial^{\mu}\phi-M^2\phi^\dagger\phi
-\frac12\mu^{2*}\phi\phi-\frac12\mu^2\phi^{\dagger}\phi^{\dagger}.
\label{bilinear_scalar_lagrangian}
\end{equation}

The non-derivative (mass) part of the Lagrangian is easily diagonalized by
a~rotation in the $\phi-\phi^{\dagger}$ space. The spectrum then contains two
real spin-0 particles with masses
$$
M_{1,2}^2=M^2\pm\left|\mu^2\right|.
$$
The corresponding real fields $\varphi_1$ and $\varphi_2$ are defined through
$$
\phi=\frac1{\sqrt2}e^{i\alpha}(\varphi_1+i\varphi_2),
$$
where the `mixing angle' $\alpha$ is merely given by the phase of the
anomalous mass term, $\tan2\alpha=\Im\mu^2/\Re\mu^2$.

\begin{figure}
$$
\def\figlab{\varphi_1}
\parbox{30\unitlength}{%
\begin{fmfgraph*}(30,15)
\fmfkeep{loop}
\fmfset{arrow_len}{0.08w}
\fmfleft{l}
\fmfright{r}
\fmf{fermion,tension=4,label=$\psi_L$,l.si=left}{vl,l}
\fmf{fermion,tension=4,label=$\psi_R$,l.si=left}{r,vr}
\fmf{dashes,right,label=$\figlab$,l.si=right}{vr,vl}
\fmf{phantom,right}{vl,vr}
\fmffreeze
\fmfipath{p[]}
\fmfiset{p1}{vpath(__vr,__vl)}
\fmfiset{p2}{vpath(__vl,__vr)}
\fmfiv{d.sh=circle,d.si=0.1w,d.fi=full}{point length(p2)/2 of p2}
\fmfi{fermion,label=$\psi_R$,l.si=left}{subpath(length(p2)/2,0) of p2}
\fmfi{fermion,label=$\psi_L$,l.si=left}{subpath(length(p2),length(p2)/2) of p2}
\end{fmfgraph*}}
\,-\,
\parbox{30\unitlength}{\def\figlab{\varphi_2}\fmfreuse{loop}}
$$
\caption{One-loop SD equation for the fermion self-energy expressed in terms of
the physical fields $\varphi_1$ and $\varphi_2$.}
\label{Fig:fermion_propagator_phi12}
\end{figure}
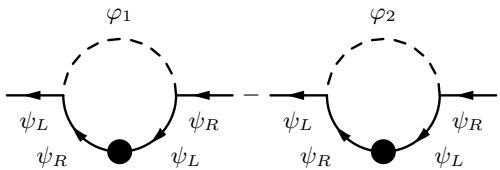

Note also that the anomalous propagator $\langle\phi\phi\rangle$ is now equal
to
$\frac12e^{2i\alpha}\left(\langle\varphi_1\varphi_1\rangle-\langle\varphi_2\varphi_2\rangle\right)$
that is, the one-loop graphs in Fig. \ref{Fig:fermion_propagator} may be
replaced with Fig. \ref{Fig:fermion_propagator_phi12}. The difference of scalar
propagators significantly improves the convergence of the Schwinger--Dyson
kernel.

\section{Formal developments}
\subsection{The scalar Nambu doublet}
We have seen in the previous section that once chiral symmetry is spontaneously
broken the scalar field $\phi$ develops non-zero anomalous propagator mixing it
with $\phi^{\dagger}$. For sake of simplicity of the general formulas to come,
we introduce a~doublet field
$$
\Phi=\left(
\begin{array}{c}
\phi \\
\phi^{\dagger}
\end{array}\right),
$$
and the matrix propagator which contains as its entries both normal and
anomalous two-point functions of the field $\phi$,
\begin{multline*}
iD(x-y)=\bra0T\{\Phi(x)\Phi^{\dagger}(y)\}\ket0=\\
=\left(
\begin{array}{cc}
\bra0T\{\phi(x)\phi^{\dagger}(y)\}\ket0 &
\bra0T\{\phi(x)\phi(y)\}\ket0\\
\bra0T\{\phi^{\dagger}(x)\phi^{\dagger}(y)\}\ket0 &
\bra0T\{\phi^{\dagger}(x)\phi(y)\}\ket0
\end{array}\right).
\end{multline*}

Note that this notation is similar to the Nambu formalism frequently used in
the theory of superconductivity \cite{Nambu:1960tm}. It is also quite natural
due to its analogy in the fermion sector. There, we could have well introduced
two `normal' propagators for the left- and right-handed chiral fields and treat
the desired mass term connecting $\psi_L$ with $\psi_R$ as an `anomalous' part
of the propagator. Instead, we work with the Dirac field
$$
\psi=\left(
\begin{array}{c}
\psi_L \\
\psi_R
\end{array}\right),
$$
which incorporates both chiral fields in a~single four-component spinor. The
chirality-changing part of the fermion propagator is then given by the
off-diagonal component of the full Dirac propagator \footnote{We work in the
chiral basis of the Dirac $\gamma$-matrices, in which $\gamma_5$ is diagonal,
and quite deliberately denote by $\psi_{L,R}$ both the two-component Weyl
spinors and the four-component Dirac spinors with just the upper two or the
lower two entries non-zero. It should be clear from the context which of these
two spinors is actually used.},
\begin{multline*}
iS(x-y)=\bra0T\{\psi(x)\bar\psi(y)\}\ket0=\\
=\left(
\begin{array}{cc}
\bra0T\{\psi_L(x)\bar\psi_L(y)\}\ket0 &
\bra0T\{\psi_L(x)\bar\psi_R(y)\}\ket0\\
\bra0T\{\psi_R(x)\bar\psi_L(y)\}\ket0 &
\bra0T\{\psi_R(x)\bar\psi_R(y)\}\ket0
\end{array}\right).
\end{multline*}
The propagators of $\psi_1$ and $\psi_2$ will be denoted as $S_{1}$ and
$S_{2}$, respectively.

\subsection{The Schwinger--Dyson equations}
\label{subsec_SDeqs} As already stressed above, spontaneous symmetry breaking
cannot be revealed at any finite order of perturbation theory. In order to deal
with this non-perturbative effect, we employ the technique of the
Schwinger--Dyson equations.

\begin{figure}
\begin{align*}
\parbox{20\unitlength}{%
\begin{fmfgraph*}(20,15)
\fmfset{arrow_len}{0.12w}
\fmfleft{l}
\fmfright{r}
\fmf{fermion}{r,v,l}
\fmfv{d.sh=circle,d.si=0.3w,d.fi=shaded}{v}
\end{fmfgraph*}}
\,&=\,
\parbox{30\unitlength}{%
\begin{fmfgraph*}(30,20)
\fmfset{arrow_len}{0.08w}
\fmfleft{l}
\fmfright{r}
\fmf{fermion,tension=3}{vl,l}
\fmf{fermion,tension=3}{r,vr}
\fmf{phantom,right}{vr,vl}
\fmf{phantom,right}{vl,vr}
\fmffreeze
\fmfipath{p[]}
\fmfiset{p1}{vpath(__vr,__vl)}
\fmfiset{p2}{vpath(__vl,__vr)}
\fmfi{dbl_dashes}{subpath(0,length(p1)/2) of p1}
\fmfi{dbl_dashes}{subpath(length(p1),length(p1)/2) of p1}
\fmfi{fermion}{subpath(length(p2)/2,0) of p2}
\fmfi{fermion}{subpath(length(p2),length(p2)/2) of p2}
\fmfiv{d.sh=circle,d.si=0.1w,d.fi=full}{point length(p1)/2 of p1}
\fmfiv{d.sh=circle,d.si=0.1w,d.fi=full}{point length(p2)/2 of p2}
\end{fmfgraph*}}\\
\parbox{20\unitlength}{%
\begin{fmfgraph*}(20,15)
\fmfset{arrow_len}{0.12w}
\fmfleft{l}
\fmfright{r}
\fmf{dbl_dashes}{r,v,l}
\fmfv{d.sh=circle,d.si=0.3w,d.fi=shaded}{v}
\end{fmfgraph*}}
\,&=\,
\parbox{30\unitlength}{%
\begin{fmfgraph*}(30,15)
\fmfset{arrow_len}{0.08w}
\fmfleft{l}
\fmfright{r}
\fmf{dbl_dashes,tension=3}{vl,l}
\fmf{dbl_dashes,tension=3}{vr,r}
\fmf{phantom,right}{vr,vl}
\fmf{phantom,right}{vl,vr}
\fmfv{label=$1,,2$,l.an=180,l.di=0.13w}{vr}
\fmffreeze
\fmfipath{p[]}
\fmfiset{p1}{vpath(__vr,__vl)}
\fmfiset{p2}{vpath(__vl,__vr)}
\fmfiv{d.sh=circle,d.si=0.1w,d.fi=full}{point length(p1)/2 of p1}
\fmfiv{d.sh=circle,d.si=0.1w,d.fi=full}{point length(p2)/2 of p2}
\fmfi{fermion}{subpath(0,length(p1)/2) of p1}
\fmfi{fermion}{subpath(length(p1)/2,length(p1)) of p1}
\fmfi{fermion}{subpath(0,length(p2)/2) of p2}
\fmfi{fermion}{subpath(length(p2)/2,length(p2)) of p2}
\end{fmfgraph*}}
\,+\,\,
\parbox{15\unitlength}{%
\begin{fmfgraph*}(15,15)
\fmfleft{l}
\fmfright{r}
\fmf{dbl_dashes,tension=0.4}{l,v,r}
\fmf{phantom,tension=0.4}{v,v}
\fmffreeze
\fmfipath{p}
\fmfiset{p}{vpath(__v,__v)}
\fmfi{dbl_dashes}{subpath(0,length(p)/2) of p}
\fmfi{dbl_dashes}{subpath(length(p)/2,length(p)) of p}
\fmfiv{d.sh=circle,d.si=0.2w,d.fi=full}{point length(p)/2 of p}
\end{fmfgraph*}}
\end{align*}
\caption{The diagrammatical representation of the one-loop Schwinger--Dyson
equations for the fermion and scalar self-energies. The first line holds for
both $\psi_1$ and $\psi_2$. The dashed blobs stand for the proper
self-energies, while the solid blobs denote the full propagators. The double
dashed line is for the Nambu $\Phi$ doublet.}
\label{Fig:SD_eqs}
\end{figure}
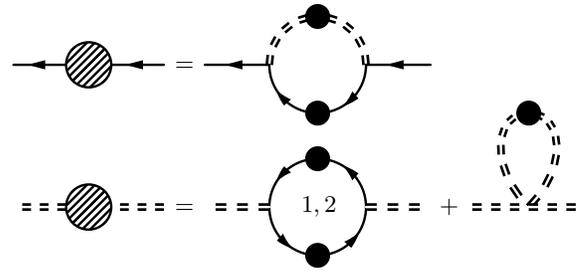

These constitute an infinite system of coupled equations for the Green's
functions of the theory. To make them more tractable, it is usual to close the
system at a~certain order by assuming a~convenient ansatz for the higher-point
Green's functions. In order to achieve a~simple gap equation we neglect all but
the two-point connected Green's functions \cite{Fetter:1971fw}. We are thus
left with a~self-consistent set of equations for the fermion and scalar
propagators, which are depicted diagrammatically in Fig. \ref{Fig:SD_eqs}. The
double-dashed line represents the Nambu doublet $\Phi$. This symbolic notation
is used to stress the fact that the Schwinger--Dyson equations for both the
symmetry-preserving and the symmetry-breaking parts of the propagators are
represented by Feynman graphs of the same topology and can thus be put in
a~simple compact form as in Fig. \ref{Fig:SD_eqs}.

We do not write down all the formulas that correspond to the Feynman diagrams
in Fig. \ref{Fig:SD_eqs}. Instead, having in mind our future simplification
neglecting all symmetry-preserving radiative corrections, we put explicitly
just the expressions for the symmetry-breaking proper self-energies. The upper
indices, $L$ or $R$ for the fermion propagators and $1$ or $2$ for the scalar
propagator, specify the matrix elements of the two-by-two matrices for $S$ and
$D$ introduced above. The same matrix notation is used for the proper
self-energies $\Sigma$ and $\Pi$.
\begin{widetext}
\begin{equation}
\begin{split}
\Sigma_1^{LR}(p)&=iy_1^2\int\frac{d^4k}{(2\pi)^4}S_1^{RL}(k)D^{12}(k-p),\\
\Sigma_2^{LR}(p)&=iy_2^2\int\frac{d^4k}{(2\pi)^4}S_2^{RL}(k)D^{21}(k-p),\\
\Pi^{12}(p)&=-iy_1^2\int\frac{d^4k}{(2\pi)^4}\Tr\bigl[S_1^{LR}(k)S_1^{LR}(k-p)\bigr]-
iy_2^2\int\frac{d^4k}{(2\pi)^4}\Tr\bigl[S_2^{RL}(k)S_2^{RL}(k-p)\bigr]+
i\lambda\int\frac{d^4k}{(2\pi)^4}D^{12}(k).
\end{split}
\label{SD_equations}
\end{equation}
\end{widetext}

Before concluding the discussion of the Schwinger--Dyson equations, let us note
that neglecting corrections to the interaction vertices is not that all
consistent with the envisaged spontaneous symmetry breaking. In Sec.
\ref{subsec_WI} we explain that the coupling of both the fermions and the
scalar to the current of the broken symmetry (that is, the axial current)
develops a~pole due to the intermediate massless Nambu--Goldstone boson state.
Now the same is also true for the (pseudo)scalar Yukawa coupling of the fermion
and scalar. As a~result, the full $\Phi$ propagator should possess a~massless
pole due to the Nambu--Goldstone boson, with the residuum determined by the
factor $\bra0\phi\ket{\pi(q)}$.

Neglecting the vertex corrections therefore means that we are not going to
reproduce correctly the whole analytical structure of the propagators. Our set
of Schwinger--Dyson equations amounts to resummation of a~certain class of
Feynman diagrams which, however, is sufficient to discover spontaneous symmetry
breaking. In other words, we are looking for spontaneous symmetry breaking in
the spectrum of the elementary excitations of the fields $\psi_{1,2}$ and
$\Phi$. The discussion of the collective excitations, which of course also
manifest themselves in the full propagators, is deferred to Sec.
\ref{subsec_WI}.

\subsection{The Ward identities}
\label{subsec_WI}
We now exploit the symmetry properties of the theory. At the
classical level, the $U(1)_{V1}\times U(1)_{V2}\times U(1)_A$ invariance of the
Lagrangian \eqref{lagrangian} implies the existence of three conserved Noether
currents --- two vector, see Eq. \eqref{Noether_vector}, and one axial, see Eq.
\eqref{Noether_axial}.

At the level of quantum field theory, the conservation of the vector and axial
currents is expressed in terms of the set of Ward identities for the Green's
functions containing the current operators. We investigate here the three-point
functions with the conserved current and a~fermion or scalar pair,
respectively.

The vector currents couple only to the fermions, so there is just one
non-trivial Ward identity for each, for the vertex functions
$G_{V1}^{\mu}(x,y,z)=\bra0T\{j_{V1}^{\mu}(x)\psi_1(y)\bar\psi_1(z)\}\ket0$ and
$G_{V2}^{\mu}(x,y,z)=\bra0T\{j_{V2}^{\mu}(x)\psi_2(y)\bar\psi_2(z)\}\ket0$,
which have the well known form
\begin{align*}
q_{\mu}\Gamma^{\mu}_{V1}(p+q,p)&=S_{1}^{-1}(p+q)-S_{1}^{-1}(p),\\
q_{\mu}\Gamma^{\mu}_{V2}(p+q,p)&=S_{2}^{-1}(p+q)-S_{2}^{-1}(p).
\end{align*}
The proper vertex functions $\Gamma^{\mu}_{V1,2}$ correspond to
$G_{V1,2}^{\mu}$ with full fermion propagators of the external legs cut off.

The axial current, on the other hand, couples to both the fermions and the
scalar. There are hence altogether three vertex functions,
$G_{A\psi_1}^{\mu}(x,y,z)=\bra0T\{j_A^{\mu}(x)\psi_1(y)\bar\psi_1(z)\}\ket0$,
$G_{A\psi_2}^{\mu}(x,y,z)=\bra0T\{j_A^{\mu}(x)\psi_2(y)\bar\psi_2(z)\}\ket0$,
and
$G_{A\phi}^{\mu}(x,y,z)=\bra0T\{j_A^{\mu}(x)\Phi(y)\Phi^{\dagger}(z)\}\ket0$.
The corresponding Ward identities read
\begin{equation}
\begin{split}
q_{\mu}\Gamma^{\mu}_{A\psi_1}(p+q,p)&=S_1^{-1}(p+q)\gamma_5+\gamma_5S_1^{-1}(p),\\
q_{\mu}\Gamma^{\mu}_{A\psi_2}(p+q,p)&=-S_2^{-1}(p+q)\gamma_5-\gamma_5S_2^{-1}(p),\\
q_{\mu}\Gamma^{\mu}_{A\phi}(p+q,p)&=-2D^{-1}(p+q)\Xi+2\Xi D^{-1}(p).
\end{split}
\label{Ward_identities}
\end{equation}
The matrix $\Xi$,
$$
\Xi=\left(\begin{array}{cr}
1 & 0\\
0 & -1
\end{array}\right),
$$
operates in the $\phi-\phi^{\dagger}$ space and is quite analogous to
$\gamma_5$ in the fermion sector.

Before closing the general discussion of the Ward identities, let us remark
that the identities of Eq. \eqref{Ward_identities} must hold whether the
symmetry is spontaneously broken or not. In both cases they strongly constrain
the form of the vertex functions, but particularly if the symmetry is broken,
they allow us to visualize the massless collective excitation predicted by the
Goldstone theorem, as we will now see.

\subsection{The Nambu--Goldstone boson}
Once chiral symmetry is broken, there must exist a~massless Nambu--Goldstone
boson in the spectrum of the theory, which couples to the Noether current of
the broken symmetry. It is the axial current which is broken in our case, and
as it couples to both the fermions and the scalar, the Nambu--Goldstone boson
must be a~collective excitation of both fermions and scalars.

General analytical properties of Green's functions imply the existence of
a~pole corresponding to an intermediate particle, once the total momentum of
a~proper subset of external legs approaches the mass shell of the particle
\cite{Weinberg:1995v1}. This means that the Nambu--Goldstone boson can be seen
as a~pole in the vertex functions of the axial current as $q^2\to0$. Its
properties are then expressed in terms of the symmetry-breaking self-energies
of the fermions and the scalar, which are obtained by solving the set of
Schwinger--Dyson equations stated in Sec. \ref{subsec_SDeqs}.

To proceed further, we approximate the proper vertex functions
$\Gamma^{\mu}_{A\psi_{1,2}}$ and $\Gamma^{\mu}_{A\phi}$ by the sum of the bare
vertex and the pole contribution. We follow the analysis of Jackiw and Johnson
\cite{Jackiw:1973tr}.

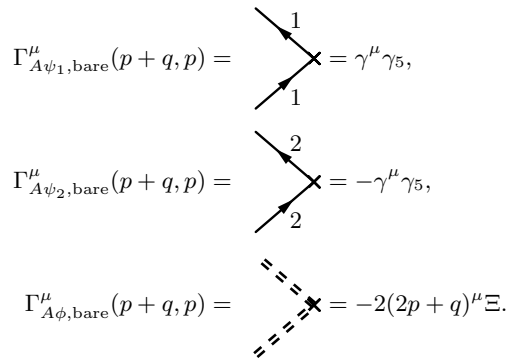
\begin{figure}
\begin{align*}
\Gamma^{\mu}_{A\psi_1\text{,bare}}(p+q,p)&=
\parbox{10\unitlength}{%
\begin{fmfgraph*}(10,15)
\fmfset{arrow_len}{0.16h}
\fmfsurroundn{v}{3}
\fmf{fermion,label=$1$,l.si=right,l.di=0.1w}{v3,v1,v2}
\fmfv{d.sh=cross,d.si=0.14h}{v1}
\end{fmfgraph*}}
\,\,=\gamma^{\mu}\gamma_5,\\
\Gamma^{\mu}_{A\psi_2\text{,bare}}(p+q,p)&=
\parbox{10\unitlength}{%
\begin{fmfgraph*}(10,15)
\fmfset{arrow_len}{0.16h}
\fmfsurroundn{v}{3}
\fmf{fermion,label=$2$,l.si=right,l.di=0.1w}{v3,v1,v2}
\fmfv{d.sh=cross,d.si=0.14h}{v1}
\end{fmfgraph*}}
\,\,=-\gamma^{\mu}\gamma_5,\\
\Gamma^{\mu}_{A\phi\text{,bare}}(p+q,p)&=
\parbox{10\unitlength}{%
\begin{fmfgraph*}(10,15)
\fmfsurroundn{v}{3}
\fmf{dbl_dashes}{v3,v1,v2}
\fmfv{d.sh=cross,d.si=0.14h}{v1}
\end{fmfgraph*}}
\,\,=-2(2p+q)^{\mu}\Xi.
\end{align*}
\caption{The bare parts of the proper vertex functions of the axial current.
The crosses indicate the axial current. The numbers at the solid lines
distinguish between $\psi_1$ and $\psi_2$.}
\label{Fig:vertex_functions_bare_part}
\end{figure}

The bare vertices are determined by the usual rules of perturbation theory. For
sake of later reference and to show how the $\Phi$ notation works, we fix them
down in Fig. \ref{Fig:vertex_functions_bare_part}.

\begin{figure}
\begin{align*}
\Gamma^{\mu}_{A\psi_1\text{,pole}}&=
\parbox{30\unitlength}{%
\begin{fmfgraph*}(30,15)
\fmfset{arrow_len}{0.08w}
\fmfleftn{l}{2}
\fmfright{r}
\fmf{fermion,tension=2,label=$1$,l.si=right,l.di=0.03w}{l1,vl,l2}
\fmf{dbl_plain,tension=2}{vl,vc}
\fmfv{d.sh=circle,d.si=0.05w,d.fi=empty}{vl}
\fmfv{d.sh=circle,d.si=0.05w,d.fi=empty}{vc}
\fmfv{d.sh=cross,d.si=0.07w,label=$1,,2$,l.an=180,l.di=0.1w}{r}
\fmf{fermion,right}{r,vc,r}
\end{fmfgraph*}}
\,\,+
\parbox{30\unitlength}{%
\begin{fmfgraph*}(30,15)
\fmfset{arrow_len}{0.08w}
\fmfleftn{l}{2}
\fmfright{r}
\fmf{fermion,tension=2,label=$1$,l.si=right,l.di=0.03w}{l1,vl,l2}
\fmf{dbl_plain,tension=2}{vl,vc}
\fmfv{d.sh=circle,d.si=0.05w,d.fi=empty}{vl}
\fmfv{d.sh=circle,d.si=0.05w,d.fi=empty}{vc}
\fmfv{d.sh=cross,d.si=0.07w}{r}
\fmf{dbl_dashes,right}{r,vc,r}
\end{fmfgraph*}}\\
\Gamma^{\mu}_{A\psi_2\text{,pole}}&=
\parbox{30\unitlength}{%
\begin{fmfgraph*}(30,15)
\fmfset{arrow_len}{0.08w}
\fmfleftn{l}{2}
\fmfright{r}
\fmf{fermion,tension=2,label=$2$,l.si=right,l.di=0.03w}{l1,vl,l2}
\fmf{dbl_plain,tension=2}{vl,vc}
\fmfv{d.sh=circle,d.si=0.05w,d.fi=empty}{vl}
\fmfv{d.sh=circle,d.si=0.05w,d.fi=empty}{vc}
\fmfv{d.sh=cross,d.si=0.07w,label=$1,,2$,l.an=180,l.di=0.1w}{r}
\fmf{fermion,right}{r,vc,r}
\end{fmfgraph*}}
\,\,+
\parbox{30\unitlength}{%
\begin{fmfgraph*}(30,15)
\fmfset{arrow_len}{0.08w}
\fmfleftn{l}{2}
\fmfright{r}
\fmf{fermion,tension=2,label=$2$,l.si=right,l.di=0.03w}{l1,vl,l2}
\fmf{dbl_plain,tension=2}{vl,vc}
\fmfv{d.sh=circle,d.si=0.05w,d.fi=empty}{vl}
\fmfv{d.sh=circle,d.si=0.05w,d.fi=empty}{vc}
\fmfv{d.sh=cross,d.si=0.07w}{r}
\fmf{dbl_dashes,right}{r,vc,r}
\end{fmfgraph*}}\\
\Gamma^{\mu}_{A\phi\text{,pole}}&=
\parbox{30\unitlength}{%
\begin{fmfgraph*}(30,15)
\fmfset{arrow_len}{0.08w}
\fmfleftn{l}{2}
\fmfright{r}
\fmf{dbl_dashes,tension=2}{l1,vl,l2}
\fmf{dbl_plain,tension=2}{vl,vc}
\fmfv{d.sh=circle,d.si=0.05w,d.fi=empty}{vl}
\fmfv{d.sh=circle,d.si=0.05w,d.fi=empty}{vc}
\fmfv{d.sh=cross,d.si=0.07w,label=$1,,2$,l.an=180,l.di=0.1w}{r}
\fmf{fermion,right}{r,vc,r}
\end{fmfgraph*}}
\,\,+
\parbox{30\unitlength}{%
\begin{fmfgraph*}(30,15)
\fmfset{arrow_len}{0.08w}
\fmfleftn{l}{2}
\fmfright{r}
\fmf{dbl_dashes,tension=2}{l1,vl,l2}
\fmf{dbl_plain,tension=2}{vl,vc}
\fmfv{d.sh=circle,d.si=0.05w,d.fi=empty}{vl}
\fmfv{d.sh=circle,d.si=0.05w,d.fi=empty}{vc}
\fmfv{d.sh=cross,d.si=0.07w}{r}
\fmf{dbl_dashes,right}{r,vc,r}
\end{fmfgraph*}}
\end{align*}
\caption{The pole parts of the proper vertex functions of the axial current.
The double solid line represents the Nambu--Goldstone boson and the empty
circles its effective vertices with the fermions and the scalar, respectively.
Both $\psi_1$ and $\psi_2$ can circulate in the closed fermion loops.}
\label{Fig:vertex_functions_pole_part}
\end{figure}
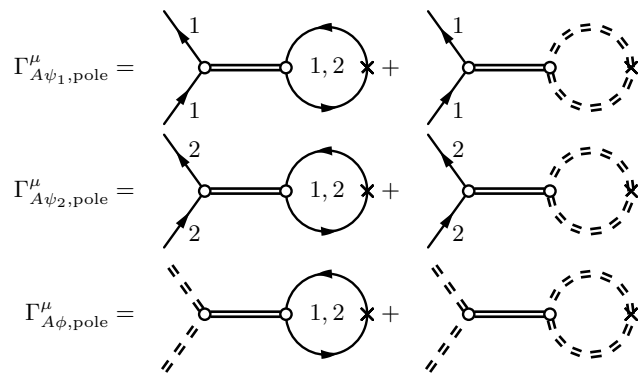

The pole of the vertex functions arises from the propagator of the intermediate
Nambu--Goldstone boson, see Fig. \ref{Fig:vertex_functions_pole_part}. The yet
unknown effective vertices of the Nambu--Goldstone boson with the fermions and
the scalar, to be extracted from the Ward identities \eqref{Ward_identities},
are denoted by empty circles in the figures and by $P_{\psi_{1,2}}(p+q,p)$ and
$P_{\phi}(p+q,p)$ in the formulas.

We can now write down the formulas for the pole contributions,
\begin{equation}
\begin{split}
\Gamma^{\mu}_{A\psi_1\text{,pole}}&=P_{\psi_1}(p+q,p)\frac i{q^2}
\left[I^{\mu}_{\psi_1}(q)+I^{\mu}_{\psi_2}(q)+I^{\mu}_{\phi}(q)\right],\\
\Gamma^{\mu}_{A\psi_2\text{,pole}}&=P_{\psi_2}(p+q,p)\frac i{q^2}
\left[I^{\mu}_{\psi_1}(q)+I^{\mu}_{\psi_2}(q)+I^{\mu}_{\phi}(q)\right],\\
\Gamma^{\mu}_{A\phi\text{,pole}}&=P_{\phi}(p+q,p)\frac i{q^2}
\left[I^{\mu}_{\psi_1}(q)+I^{\mu}_{\psi_2}(q)+I^{\mu}_{\phi}(q)\right],
\end{split}
\label{pole_part_integrals}
\end{equation}
where $I^{\mu}_{\psi_{1,2}}(q)$ and $I^{\mu}_{\phi}(q)$ represent the fermion
and scalar loops in Fig. \ref{Fig:vertex_functions_pole_part}. They are given
by the equations
\begin{widetext}
\begin{equation}
\begin{split}
I^{\mu}_{\psi_1}(q)&=-\int\frac{d^4k}{(2\pi)^4}\Tr\Bigl[\gamma^{\mu}\gamma_5iS_1(k-q)
P_{\psi_1}(k-q,k)iS_1(k)\Bigr],\\
I^{\mu}_{\psi_2}(q)&=-\int\frac{d^4k}{(2\pi)^4}\Tr\Bigl[-\gamma^{\mu}\gamma_5iS_2(k-q)
P_{\psi_2}(k-q,k)iS_2(k)\Bigr],\\
I^{\mu}_{\phi}(q)&=\frac12\int\frac{d^4k}{(2\pi)^4}\Tr\Bigl[-2(2k-q)^{\mu}\Xi
iD(k-q)P_{\phi}(k-q,k)iD(k)\Bigr].
\end{split}
\label{loop_integrals}
\end{equation}
\end{widetext}
The difference of the integral prefactors here arises from the different nature
of the degrees of freedom circulating in the loops. The $-1$ is the standard
fermion loop factor, while the $\frac12$ follows from the fact that by
introducing $\Phi$ instead of $\phi$ we have effectively doubled the number of
degrees of freedom, which must be compensated for when calculating the trace
over a~closed loop.

As the loop integrals $I^{\mu}_{\psi_{1,2}}(q)$ and $I^{\mu}_{\phi}(q)$ depend only
on a~single external momentum $q$, they are forced by Lorentz covariance to
have the form
\begin{align*}
I^{\mu}_{\psi_{1,2}}(q)&=-iq^{\mu}I_{\psi_{1,2}}(q^2),\\
I^{\mu}_{\phi}(q)&=-iq^{\mu}I_{\phi}(q^2).
\end{align*}
Inserting this into Eq. \eqref{pole_part_integrals} and the sum of the bare and
the pole contributions into the Ward identities \eqref{Ward_identities}, and
going onto the Nambu--Goldstone boson mass shell, $q^2\to0$, we arrive at the
general formula for the effective vertices,
\begin{widetext}
\begin{equation}
\begin{split}
P_{\psi_1}(p+q,p)&=\frac1N\left[S_1^{-1}(p+q)\gamma_5+\gamma_5S_1^{-1}(p)-\slashed
q\gamma_5\right],\\
P_{\psi_2}(p+q,p)&=-\frac1N\left[S_2^{-1}(p+q)\gamma_5+\gamma_5S_2^{-1}(p)-\slashed
q\gamma_5\right],\\
P_{\phi}(p+q,p)&=-\frac2N\left[D^{-1}(p+q)\Xi-\Xi D^{-1}(p)-q\cdot(2p+q)\Xi\right],
\end{split}
\label{effective_vertices_general}
\end{equation}
\end{widetext}
where the normalization factor $N$ is given by $N=
I_{\psi_1}(0)+I_{\psi_2}(0)+I_{\phi}(0)$. As noted by Jackiw and Johnson
\cite{Jackiw:1973tr}, the effective vertices found this way are ambiguous at
the order $O(q^2)$, since we have approximated the loop corrections to the full
proper vertex functions $\Gamma^{\mu}_{A\psi_{1,2}}$ and $\Gamma^{\mu}_{A\phi}$
by their pole parts, and completely neglected other finite contributions.

To complete the calculation of the vertex functions, we must now plug the
expressions \eqref{effective_vertices_general} back into Eq.
\eqref{loop_integrals} and solve the resulting system of equations for
$I_{\psi_{1,2}}(0)$ and $I_{\phi}(0)$.

\section{Model results}
\subsection{Effective vertices and loop integrals}
The results obtained above are fairly general as we have used only very few and
weak assumptions like the pole term dominance in the proper vertex functions.
On the other hand, it is not easy to draw any concrete results from formulas
like Eqs. \eqref{loop_integrals} and \eqref{effective_vertices_general}. To
push our conclusions little bit further, we now make a~simplifying assumption
that will allow us to finish the calculation.

Since we are looking for spontaneous symmetry breaking, we shall neglect
ordinary (symmetry-preserving) renormalization of the fermion and scalar
propagators \cite{Cornwall:1974vz} and retain just the symmetry-breaking
self-energies. This will enable us to proceed analytically as far as possible.
Of course, as soon as one pretends at phenomenological relevance of the
obtained results, all radiative corrections must be included, but this is not
the aim of the present paper. Here we just wish to \emph{demonstrate} that
spontaneous symmetry breaking is possible in a~model like Eq.
\eqref{lagrangian}.

We thus make the following ansatz for the fermion and scalar propagators,
\begin{equation}
\begin{split}
S_{1,2}^{-1}(p)&=\slashed p-\Sigma_{1,2}(p),\\
D^{-1}(p)&=\left(
\begin{array}{cc}
p^2-M^2 & -\Pi(p) \\
-\Pi^*(p) & p^2-M^2
\end{array}\right),
\end{split}
\label{ansatz_propagators}
\end{equation}
where $\Sigma_{1,2}(p)$ are the Lorentz-scalar chirality-changing proper
self-energies, and $\Pi(p)$ is the anomalous proper self-energy of the scalar
field.

With this assumption, the effective vertices \eqref{effective_vertices_general}
become
\begin{equation}
\begin{split}
P_{\psi_1}(p+q,p)&=-\frac1N\left[\Sigma_1(p+q)+\Sigma_1(p)\right]\gamma_5,\\
P_{\psi_2}(p+q,p)&=\frac1N\left[\Sigma_2(p+q)+\Sigma_2(p)\right]\gamma_5,\\
P_{\phi}(p+q,p)&=-\frac2N\left(
\begin{array}{cc}
0 & \Pi(p+q)+\Pi(p) \\
-\Pi^*(p+q)-\Pi^*(p) & 0
\end{array}\right).
\label{effective_vertices}
\end{split}
\end{equation}

We can now go on to evaluate the last missing piece that is, the normalization
factors $I_{\psi_{1,2}}(0)$ and $I_{\phi}(0)$. We substitute for the
propagators \eqref{ansatz_propagators} and the effective vertices
\eqref{effective_vertices} in the loop integrals \eqref{loop_integrals}, which
turn out to be parameterized in terms of the integrals (for sake of readability
we put the arguments of $\Sigma_{1,2}$ and $\Pi$ to the lower index),
\begin{widetext}
\begin{equation}
\begin{split}
-iq^{\mu}J_{\psi_1}(q^2)&=8\int\frac{d^4k}{(2\pi)^4}\frac{(k-q)^{\mu}\Sigma_{1,k}}{k^2-\Sigma^2_{1,k}}
\frac{\Sigma_{1,k}+\Sigma_{1,k-q}}{(k-q)^2-\Sigma^2_{1,k-q}},\\
-iq^{\mu}J_{\psi_2}(q^2)&=8\int\frac{d^4k}{(2\pi)^4}\frac{(k-q)^{\mu}\Sigma_{2,k}}{k^2-\Sigma^2_{2,k}}
\frac{\Sigma_{2,k}+\Sigma_{2,k-q}}{(k-q)^2-\Sigma^2_{2,k-q}},\\
-iq^{\mu}J_{\phi}(q^2)&=8\int\frac{d^4k}{(2\pi)^4}\frac{(2k-q)^{\mu}(k^2-M^2)}{(k^2-M^2)^2-|\Pi_k|^2}
\frac{\Re\left[\Pi^*_{k-q}\bigl(\Pi_{k}+\Pi_{k-q}\bigr)\right]}{\bigl[(k-q)^2-M^2\bigr]^2-|\Pi_{k-q}|^2}.
\end{split}
\label{J_integrals}
\end{equation}
\end{widetext}
With these definitions, the expressions for $I_{\psi_{1,2}}(q^2)$ and
$I_{\phi}(q^2)$ read
\begin{equation*}
\begin{split}
I_{\psi_{1,2}}(q^2)&=\frac{J_{\psi_{1,2}}(q^2)}N,\\
I_{\phi}(q^2)&=\frac{J_{\phi}(q^2)}N,
\end{split}
\end{equation*}
wherefrom the overall normalization factor of the vertices in Eq.
\eqref{effective_vertices} is equal to
\begin{equation}
N=\sqrt{J_{\psi_1}(0)+J_{\psi_2}(0)+J_{\phi}(0)}.
\end{equation}

Let us finally note that the ansatz \eqref{ansatz_propagators} also allows us
to simplify the Schwinger--Dyson equations \eqref{SD_equations}, whose solution
will be the subject of the next section. With the same index notation as above
in Eq. \eqref{J_integrals}, Eqs. \eqref{SD_equations} become
\begin{widetext}
\begin{equation}
\begin{split}
\Sigma_{1,p}&=iy_1^2\int\frac{d^4k}{(2\pi)^4}\frac{\Sigma_{1,k}}{k^2-\Sigma_{1,k}^2}
\frac{\Pi_{k-p}}{\left[(k-p)^2-M^2\right]^2-|\Pi_{k-p}|^2},\\
\Sigma_{2,p}&=iy_2^2\int\frac{d^4k}{(2\pi)^4}\frac{\Sigma_{2,k}}{k^2-\Sigma_{2,k}^2}
\frac{\Pi^*_{k-p}}{\left[(k-p)^2-M^2\right]^2-|\Pi_{k-p}|^2},\\
\Pi_p&=-\sum_{j=1,2}2iy_j^2\int\frac{d^4k}{(2\pi)^4}\frac{\Sigma_{j,k}}{k^2-\Sigma_{j,k}^2}
\frac{\Sigma_{j,k-p}}{(k-p)^2-\Sigma_{j,k-p}^2}+i\lambda\int\frac{d^4k}{(2\pi)^4}
\frac{\Pi_k}{(k^2-M^2)^2-|\Pi_k|^2}.
\end{split}
\label{SD_model_equations}
\end{equation}
\end{widetext}

\subsection{Solution to the Schwinger--Dyson equations}
We now wish to demonstrate that the Schwinger--Dyson equations
\eqref{SD_model_equations} actually do have a~non-trivial solution that is, our
mechanism is capable of generating fermion masses dynamically. To that end note
that Eqs. \eqref{SD_model_equations} constitute a~set of coupled non-linear
integral equations for the unknown functions $\Sigma_{1,p}$, $\Sigma_{2,p}$ and
$\Pi_p$. This is still too complicated to deal with, and we therefore introduce
further simplifications. We keep in mind that we are just attempting to break
the chiral symmetry dynamically, and do not intend to produce any
phenomenological conclusions at this stage.

First, we abandon the $\lambda$-term in the last of Eqs.
\eqref{SD_model_equations}. Formally, the advantage of this step is in the fact
that $\Pi$ is then expressed exclusively in terms of the $\Sigma$'s.
Physically, the symmetry is broken by the strong dynamics of the Yukawa
interaction, while the scalar field quartic self-interaction can be switched on
perturbatively later. This possibility is in marked contrast with the standard
use of a~condensing scalar. The $\lambda(\phi^{\dagger}\phi)$ term in the
Lagrangian will become indispensable as a~counter-term after including
ordinary, symmetry-preserving quantum corrections.

Second, we consider for simplicity just the special case $y_1=y_2$. The
Lagrangian is then invariant under the discrete symmetry
$\psi_1\leftrightarrow\psi_2$, $\phi\leftrightarrow\phi^{\dagger}$. As far as
the induced anomalous scalar self-energy is real, we may assume that this
discrete symmetry is not spontaneously broken and the self-energies $\Sigma_1$
and $\Sigma_2$ are therefore equal. We are thus left with two coupled equations
for $\Pi$ and just one $\Sigma$.

The reduced set of equations may now in principle be solved iteratively. We
performed a~numerical calculation to estimate the order of magnitude of the
generated fermion mass. We used the Euclidean approximation that is, made
a~formal Wick rotation of the momenta in Eqs. \eqref{SD_model_equations}. We
made use of the fact that, after canceling the $\lambda$-term, the right hand
side of the last of Eqs. \eqref{SD_model_equations} depends just on the
$\Sigma$'s.

\begin{figure}
\begin{center}
\framebox{\scalebox{0.75}{\include{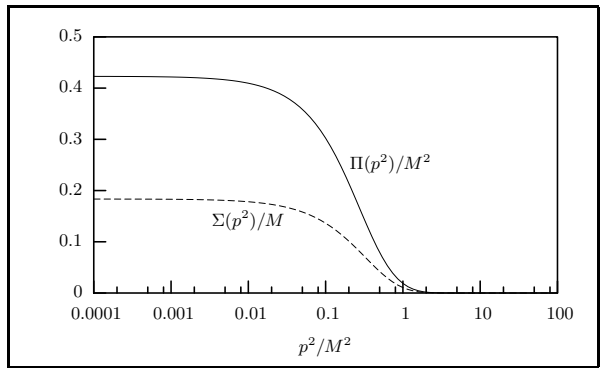}}}
\end{center}
\caption{Results of the numerical computation of the self-energies $\Sigma$ and
$\Pi$.}
\label{Fig:num_results}
\end{figure}

We took an initial ansatz for the $\Sigma$ and used it to calculate the zeroth
approximation for the $\Pi$. After then, we solved the two coupled equations
iteratively. Our results are summarized by the graphs in Fig.
\ref{Fig:num_results}.

We can see that the Eqs. \eqref{SD_model_equations} indeed possess
a~non-trivial solution. As far as we were able to check, this solution is
unique in the sense that it is independent of the initial ansatz for the
$\Sigma's$. The self-energies fall down rapidly once the momentum exceeds the
bare scalar mass $M$, thus verifying our assumptions on the convergence of the
loop integrals.

It would be perhaps more appropriate to do all the calculations in the
Minkowski space as the physical mass lies in the time-like region of momenta.
Such a~calculation has been performed in Ref. \cite{Bicudo:2003fd}.

\section{Discussion and conclusion}
Within simplifying assumptions stated in the text we have demonstrated that
a~strong chirally invariant Yukawa interaction of massless fermions with
a~massive complex scalar field can generate the fermion masses by genuinely
quantum (i.e. non-perturbative) loop effects. By the existence theorem this
implies the massless pseudoscalar Nambu--Goldstone boson in the spectrum. In
this respect our program is very much the same as that of the renown NJL paper
\cite{Nambu:1961tp}.

We believe that certain appeal of our suggestion is in the ultraviolet
finiteness of non-perturbatively calculated quantities. It can be traced to the
necessity of a~generic coupling of the Schwinger--Dyson equations for the
fermion and the scalar field propagators. It is definitely more subtle that the
single Schwinger--Dyson equation for the fermion propagator with chirality
conserving vector interactions. Technically the $\phi-\phi^{\dagger}$ mixing
results in the difference of propagators of the scalar mass eigenstates and,
consequently, in decent ultraviolet behavior of anomalous (symmetry-breaking)
loop integrals.

Vexing assumptions of the present exploratory stage of the development of the
model have to be replaced by better ones. In particular it is desirable to have
approximate analytic solutions $\Sigma(p^2)$ and $\Pi(p^2)$ in Minkowski space.
The fermion masses $m_{1,2}$ are determined by solving
$m_{1,2}^2=\Sigma^2_{1,2}(p^2=m_{1,2}^2)$ and by the dimensional argument the
solution must have the form
\begin{equation}
m_{1,2}=Mf_{1,2}(y_{1,2}).
\label{mass_formula}
\end{equation}
Preliminary numerical analysis in Euclidean space suggests that
$\Sigma_{1,2}\to0$ for $y_{1,2}$ approaching a~(large) critical value. Our
numerical calculation gives a~rough estimate $y_{\text{crit.}}\approx35$. The
formula \eqref{mass_formula} is to be compared with
$m_{1,2}=\frac1{\sqrt2}vy_{1,2}$ of the standard tree-level approach with
a~condensing scalar.

Possible uses of our model, if harshly justified, are also those of NJL in its
contemporary interpretation:

(i) The apparently non-BCS-like form of the Schwinger--Dyson equations for the
fermion propagator \eqref{SD_equations} is suggestive for modeling fermionic
superfluidity with scalar effective degrees of freedom.

(ii) Non-Abelian generalization and gauging of the NJL model resulted in the
past in models of the dynamical mass generation in $SU(2)_L\times U(1)_Y$
gauge-invariant electroweak models \cite{Hosek:1985jr,Bardeen:1990ds}. With our
way of treating scalars we plan to follow the same path \cite{Brauner:2004kg}.

\begin{acknowledgments}
We are grateful to Petr Bene\v{s} for doing the numerical calculation of the
self-energies. This work was supported in part by the Institutional Research
Plan AV0Z10480505, and by the GACR doctoral project No. 202/05/H003.
\end{acknowledgments}

\end{fmffile}

\begin{thebibliography}{11}
\expandafter\ifx\csname natexlab\endcsname\relax\def\natexlab#1{#1}\fi
\expandafter\ifx\csname bibnamefont\endcsname\relax
  \def\bibnamefont#1{#1}\fi
\expandafter\ifx\csname bibfnamefont\endcsname\relax
  \def\bibfnamefont#1{#1}\fi
\expandafter\ifx\csname citenamefont\endcsname\relax
  \def\citenamefont#1{#1}\fi
\expandafter\ifx\csname url\endcsname\relax
  \def\url#1{\texttt{#1}}\fi
\expandafter\ifx\csname urlprefix\endcsname\relax\def\urlprefix{URL }\fi
\providecommand{\bibinfo}[2]{#2}
\providecommand{\eprint}[2][]{\url{#2}}

\bibitem[{\citenamefont{Quigg}(2005)}]{Quigg:2005hr}
\bibinfo{author}{\bibfnamefont{C.}~\bibnamefont{Quigg}} (\bibinfo{year}{2005}),
  \eprint{hep-ph/0502070}.

\bibitem[{\citenamefont{Nambu}(1960)}]{Nambu:1960tm}
\bibinfo{author}{\bibfnamefont{Y.}~\bibnamefont{Nambu}},
  \bibinfo{journal}{Phys. Rev.} \textbf{\bibinfo{volume}{117}},
  \bibinfo{pages}{648} (\bibinfo{year}{1960}).

\bibitem[{\citenamefont{Fetter and Walecka}(1971)}]{Fetter:1971fw}
\bibinfo{author}{\bibfnamefont{A.~L.} \bibnamefont{Fetter}} \bibnamefont{and}
  \bibinfo{author}{\bibfnamefont{J.~D.} \bibnamefont{Walecka}},
  \emph{\bibinfo{title}{Quantum theory of many-particle systems}},
  International series in pure and applied physics
  (\bibinfo{publisher}{McGraw--Hill}, \bibinfo{address}{New York},
  \bibinfo{year}{1971}).

\bibitem[{\citenamefont{Weinberg}(1995)}]{Weinberg:1995v1}
\bibinfo{author}{\bibfnamefont{S.}~\bibnamefont{Weinberg}},
  \emph{\bibinfo{title}{The Quantum Theory of Fields}},
  vol.~\bibinfo{volume}{1} (\bibinfo{publisher}{Cambridge University Press},
  \bibinfo{address}{Cambridge}, \bibinfo{year}{1995}), \bibinfo{edition}{1st}
  ed.

\bibitem[{\citenamefont{Jackiw and Johnson}(1973)}]{Jackiw:1973tr}
\bibinfo{author}{\bibfnamefont{R.}~\bibnamefont{Jackiw}} \bibnamefont{and}
  \bibinfo{author}{\bibfnamefont{K.}~\bibnamefont{Johnson}},
  \bibinfo{journal}{Phys. Rev.} \textbf{\bibinfo{volume}{D8}},
  \bibinfo{pages}{2386} (\bibinfo{year}{1973}).

\bibitem[{\citenamefont{Cornwall et~al.}(1974)\citenamefont{Cornwall, Jackiw,
  and Tomboulis}}]{Cornwall:1974vz}
\bibinfo{author}{\bibfnamefont{J.~M.} \bibnamefont{Cornwall}},
  \bibinfo{author}{\bibfnamefont{R.}~\bibnamefont{Jackiw}}, \bibnamefont{and}
  \bibinfo{author}{\bibfnamefont{E.}~\bibnamefont{Tomboulis}},
  \bibinfo{journal}{Phys. Rev.} \textbf{\bibinfo{volume}{D10}},
  \bibinfo{pages}{2428} (\bibinfo{year}{1974}).

\bibitem[{\citenamefont{Bicudo}(2004)}]{Bicudo:2003fd}
\bibinfo{author}{\bibfnamefont{P.}~\bibnamefont{Bicudo}},
  \bibinfo{journal}{Phys. Rev.} \textbf{\bibinfo{volume}{D69}},
  \bibinfo{pages}{074003} (\bibinfo{year}{2004}), \eprint{hep-ph/0312373}.

\bibitem[{\citenamefont{Nambu and Jona-Lasinio}(1961)}]{Nambu:1961tp}
\bibinfo{author}{\bibfnamefont{Y.}~\bibnamefont{Nambu}} \bibnamefont{and}
  \bibinfo{author}{\bibfnamefont{G.}~\bibnamefont{Jona-Lasinio}},
  \bibinfo{journal}{Phys. Rev.} \textbf{\bibinfo{volume}{122}},
  \bibinfo{pages}{345} (\bibinfo{year}{1961}).

\bibitem[{\citenamefont{{Ho\v{s}ek}}(1985)}]{Hosek:1985jr}
\bibinfo{author}{\bibfnamefont{J.}~\bibnamefont{{Ho\v{s}ek}}}
  (\bibinfo{year}{1985}), \bibinfo{note}{{CERN-TH-4104/85}}.

\bibitem[{\citenamefont{Bardeen et~al.}(1990)\citenamefont{Bardeen, Hill, and
  Lindner}}]{Bardeen:1990ds}
\bibinfo{author}{\bibfnamefont{W.~A.} \bibnamefont{Bardeen}},
  \bibinfo{author}{\bibfnamefont{C.~T.} \bibnamefont{Hill}}, \bibnamefont{and}
  \bibinfo{author}{\bibfnamefont{M.}~\bibnamefont{Lindner}},
  \bibinfo{journal}{Phys. Rev.} \textbf{\bibinfo{volume}{D41}},
  \bibinfo{pages}{1647} (\bibinfo{year}{1990}).

\bibitem[{\citenamefont{Brauner and {Ho\v{s}ek}}(2004)}]{Brauner:2004kg}
\bibinfo{author}{\bibfnamefont{T.}~\bibnamefont{Brauner}} \bibnamefont{and}
  \bibinfo{author}{\bibfnamefont{J.}~\bibnamefont{{Ho\v{s}ek}}}
  (\bibinfo{year}{2004}), \eprint{hep-ph/0407339}.
\end{thebibliography}
\end{document}